\documentclass[a4paper,reprint,showkeys,nofootinbib,twoside]{revtex4-1}

\usepackage[english]{babel}
\usepackage{amsmath}
\usepackage{amssymb}
\usepackage{bbm}
\usepackage{graphicx}
\usepackage{mathtools}

\usepackage[colorinlistoftodos, color=green!40, prependcaption]{todonotes}
\usepackage{amsthm}
\usepackage{mathtools}
\usepackage{physics}
\usepackage{xcolor}
\usepackage{graphicx}
\usepackage[left=18mm,right=18mm,top=35mm,columnsep=15pt]{geometry}
\usepackage{adjustbox}
\usepackage{placeins}
\usepackage[T1]{fontenc}
\usepackage{lipsum}
\usepackage{csquotes}

\usepackage[colorlinks=true,linkcolor=blue,citecolor=blue,urlcolor=blue,unicode=true,pdftitle={Article},pdfauthor={Author}]{hyperref}

\begin{document}

\title{Feshbach Projection Formalism for Resonant Electrodynamics}

\author{Nikolai A. Vlasov$^{1}$, Ekaterina E. Maslova$^{1}$, Andrey Novitsky$^{2}$, and Mikhail V. Rybin$^{1,3}$}
\email[Correspondence email address: \\]{m.rybin@metalab.ifmo.ru}
\affiliation{$^1$School of Physics and Engineering, ITMO University, 197101, St. Petersburg, Russia\\
$^2$Belarusian State University, 220030, Minsk, Belarus\\
$^3$Ioffe Institute, 194021, St. Petersburg, Russia\\
}

\date{\today}

\begin{abstract}
Open resonant photonic systems are widely described by effective non-Hermitian formalisms, such as temporal coupled-mode theory, which capture radiative decay through complex eigenfrequencies. In conventional formulations, however, these effective models remain largely phenomenological, lacking a direct microscopic connection to the underlying Maxwell equations. To bridge this gap, we present an analytically tractable light–matter framework based on the Feshbach projection formalism applied to classical electrodynamics. Starting from a coupled system of mechanical oscillator and a one-dimensional electromagnetic field, we construct a strictly Hermitian first-order Hamiltonian. We demonstrate that rigorously eliminating the electromagnetic degrees of freedom via the outgoing Green's function automatically generates a non-Hermitian mechanical operator. Crucially, the resulting radiative damping is determined by the local electromagnetic response rather than introduced ad hoc. Furthermore, we extend this framework to two spatially separated dipoles and show how field retardation transforms a finite set of discrete modes into a countably infinite Fabry–P\'erot-like spectrum as the separation deviates from zero. The proposed framework provides an intuitive yet rigorous foundation for deriving effective non-Hermitian dynamics directly from conservative light–matter interactions.

\end{abstract}


\maketitle

\section{Introduction}

Classical electrodynamics relies on Maxwell's equations to describe the behavior of electric and magnetic fields. Typically, these equations are solved by eliminating one of the field vectors to derive a wave equation containing a second-order time derivative. To date, numerous analytical and numerical methods were proposed to solve these equations~\cite{Teixeira2023, Tsitsas2025, Erricolo2025, Ana2025, Amirjani2021, Evlyukhin2023, Pantyukhina2025}.  They enable solutions for an almost arbitrary spatial distribution of dielectric permittivity, while recent advancements in metaphotonics focus on systems comprising elements with
a pronounced resonant response~\cite{Kuznetsov2016, Kruk2017, Rybin2024, KOSHELEV2019}. Although such systems can be simulated with decent accuracy, the underlying physics often remains obscured. This is because resonant modes cannot be easily isolated from the total electromagnetic field, as they are treated as an intrinsic part of the field itself.

A systematic approach to describing resonances was developed in quantum scattering theory within the framework of the Feshbach projection formalism~\cite{FESHBACH1958, FESHBACH1962, Feshbach1964, Rotter2001, Rotter2009}. Designed to solve the time-dependent Schr\"odinger equation, which contains a first-order time derivative, this formalism partitions the state space into a subsystem of interest and its environment. In doing so, it replaces the full evolution problem with an effective equation acting in a reduced space. Eliminating the environmental degrees of freedom introduces a self-energy term that accounts for their influence on the retained subsystem. For a subsystem coupled to a continuum of scattering states, the outgoing-wave boundary condition gives rise to a non-Hermitian contribution to the effective operator, thereby directly linking resonance decay to the coupling between discrete degrees of freedom and the continuum.

In photonics, temporal coupled-mode theory (TCMT) provides a widely used reduced description of resonant dynamics~\cite{Fan2003, Fan2004, Fan2010, Fan2012, Maksimov2025}. Its equations express the evolution of resonance amplitudes and their coupling to incoming and outgoing radiation channels in a form closely related to effective Hamiltonian formalisms in quantum mechanics. In its conventional phenomenological formulation, however, this approach does not inherently establish a microscopic connection between the coupling coefficients and the underlying Maxwell equations. Some attempts have been made to establish a rigorous connection between Maxwell's equations and coupled-mode descriptions, particularly through quasinormal-mode expansions~\cite{Wu2024}. Although this approach has provided rigorous generalized description of electromagnetic resonances, it does not, in general, recover the conventional TCMT equations exactly. Moreover, while non-Hermitian operators naturally arise in quasinormal-mode descriptions of open electromagnetic systems, this formulation does not explicitly trace their emergence from an underlying Hermitian light–matter dynamics. To bridge this gap, an explicit analytical model remains highly demanded and valuable for tracing how conservative light–matter dynamics gives rise to radiative damping and a non-Hermitian resonant response.


In this work, we demonstrate how an electromagnetic system featuring resonances can be rigorously described within the Feshbach projection formalism. First, we introduce a one-dimensional (1D) dipole and describe its interaction with the electric field. Starting from the coupled Maxwell and mechanical equations, we define an energy-normalized state vector to formulate the evolution of the complete light–matter system as a Schr\"odinger-type equation. The resulting full Hamiltonian is Hermitian with respect to the energy inner product, thereby providing a conservative foundation for the reduced resonance description. We then eliminate the electromagnetic degrees of freedom using the outgoing Green's function, implementing a subsystem reduction in the spirit of the Feshbach formalism. This procedure yields an explicit effective Hamiltonian for the mechanical oscillator and directly identifies its non-Hermitian contribution with radiative self-action. Crucially, the damping term is determined by the imaginary part of the electromagnetic Green's function evaluated at the dipole position, rather than being introduced phenomenologically. We subsequently derive the scattering matrix and relate its resonance poles to the complex eigenfrequencies of the effective Hamiltonian. Finally, we extend the model to two spatially separated dipoles. In addition to the self-action of each oscillator, the reduced description reveals a field-mediated interaction characterized by a frequency-dependent propagation phase. This extension links the same light–matter formulation to the formation of Fabry–P\'erot-like resonances. Taken together, these results provide an analytically tractable connection between Hermitian full-system dynamics, non-Hermitian subsystem evolution, and resonant electromagnetic scattering.

\section{Model of 1D dipole}

\begin{figure}
	\centering
	\includegraphics[width=\columnwidth]{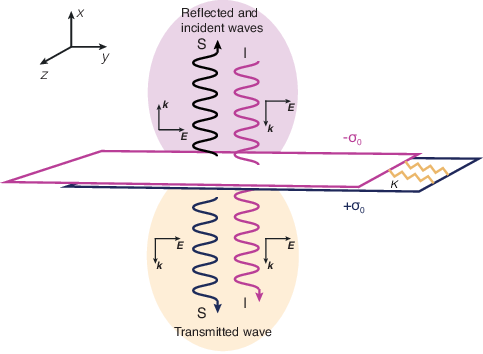}
\caption{Schematic illustration of the model configuration. The magenta rectangle represents an infinite plate in the $YZ$ plane with a negative charge density $-\sigma_0$, while the blue rectangle corresponds to an infinite plate in the $YZ$ plane with a positive charge density $+\sigma_0$. The distance between the two plates is assumed to be negligibly small. Due to an external field, the charged plates oscillate in antiphase along the $Y$-axis, radiating electromagnetic waves, which constitute the scattered wave. The yellow zigzags illustrate an elastic restoring force with coefficient $\kappa$. The labels $I$ and $S$ denote the incident and scattered waves, respectively. $\mathbf{k}$ and $\mathbf{E}$ show the directions of the wave vector and the electric field for each wave. The magenta ellipse (upper half-space) highlights the superposition of the incident and scattered waves. The yellow ellipse (lower half-space) highlights the transmitted wave, which arises from the interference of the incident wave propagating past the plates (magenta arrow) and the wave scattered into this half-space (blue arrow).}
\label{FIG:1}
\end{figure}

We begin by defining the model configuration. Consider two infinite metallic plates located in the $YZ$-plane at a vanishingly small distance from each other (see Fig.~\ref{FIG:1}), possessing opposite surface charge densities, $+\sigma_0$ and $-\sigma_0$. An incident electromagnetic field $\mathbf{E}=E\mathbf{e}_y$ propagates along the $X$-axis, normal to the plates. Under its action, the charged plates oscillate in opposite directions along the $Y$-axis, inducing a surface current and giving rise to an oscillating dipole moment per unit area. The induced current $\mathbf{j}$ radiates a scattered field into both half-spaces. Consequently, the total field in the system is composed of the incident and scattered fields on both sides of the plates. Whereas in the backward half-space, the scattered field forms the reflected wave, while in the forward half-space, its superposition with the incident field yields the transmitted wave.

To model the relative motion, two plates possess an effective surface mass density $m$ and is subject to a restoring force characterized by the coefficient $\kappa$. The relative displacement and the corresponding momentum density due to the oscillatory motion are defined as $\mathbf{u} = \mathbf{e}_y u$ and $\mathbf{p} = m \mathrm{d}\mathbf{u}/ \mathrm{d} t$, respectively. The resulting induced current density takes the form $\mathbf{j} = \mathbf{e}_y \delta (x) p\sigma/m$, where the $\delta(x)$-function accounts for the position of the plates at $x=0$.

Then, the mechanical equations projected onto the $Y$-axis, which describe the oscillations of the moving plate, are given by
\begin{equation}
\label{mechanical eqs}
\begin{cases}
    \dfrac{\mathrm{d} u}{\mathrm{d} t} = \dfrac{1}{m} p, \\
    \dfrac{\mathrm{d} p}{\mathrm{d} t} = - \kappa u + \sigma E,
\end{cases}
\end{equation}
where $\sigma = 2 \sigma_0$ denotes the effective charge per unit area associated with the relative coordinate and $E$ denotes the electric field acting on the charges. Here, the term $-\kappa u$ is the restoring force that returns the plates to equilibrium, and the term $\sigma E$ is the driving force exerted by the total electric field at the position of the plates.

Next, we analyze the electromagnetic scattering by the proposed model of a 1D dipole and incorporate Maxwell’s equations. Considering the polarization state defined by $\mathbf{E} = E \mathbf{e}_y$ and $\mathbf{H} = H \mathbf{e}_z$, Maxwell’s equations can be written as
\begin{equation}
\label{maxwell eqs}
\begin{cases}
    \dfrac{\partial E}{\partial x} = -\mu_0 \dfrac{\partial H}{\partial t}, \\
    \dfrac{\partial H}{\partial x} = -j - \varepsilon_0 \dfrac{\partial E}{\partial t}.
\end{cases}
\end{equation}
In these equations the current $j$ plays the role of the source of the scattered field.

Thus, Eqs.~\eqref{mechanical eqs} and \eqref{maxwell eqs} form a closed system of equations for the problem under consideration. By solving them simultaneously, one can obtain the full solution to the corresponding scattering problem.

\section{Hermitian light-matter dynamics}

Having established the four underlying equations required to solve the scattering problem, we note that the state of the system is fully characterized by four variables: $E, H, p$, and $u$. Since the fields $E$ and $H$ belong to the space $L^2(\mathbb{R})$, while the relative coordinate $u$ and momentum density $p$ belong to $\mathbb{C}$, the state space of the complete system is given by $L^2(\mathbb{R}) \oplus L^2(\mathbb{R}) \oplus \mathbb{C} \oplus \mathbb{C}$. It is worth noting that, strictly speaking, plane waves, which are commonly used in scattering problems, do not belong to $L^2(\mathbb{R})$, as their total energy is infinite. To address this issue, we implicitly introduce a regularizing factor that ensures their decay at spatial infinity. For simplicity, we omit this factor throughout the derivation, assuming that our analysis is restricted to a spatial region where its contribution is negligible. To equip this space with a Hilbert space structure, we must define an inner product. Let us consider two states of the system
\begin{equation}
\label{state1}
    \ket {\psi_1} = \begin{pmatrix}
        \sqrt{\dfrac{\varepsilon_0}{2}} E_1 &
        \sqrt{\dfrac{\mu_0}{2}} H_1 &
        \dfrac{1}{\sqrt{2m}}p_1 &
        \sqrt{\dfrac{\kappa}{2}}u_1
    \end{pmatrix}^T
\end{equation}
and

\begin{equation}
\label{state2}
    \ket {\psi_2} = \begin{pmatrix}
        \sqrt{\dfrac{\varepsilon_0}{2}}E_2 &
        \sqrt{\dfrac{\mu_0}{2}} H_2 &
        \dfrac{1}{\sqrt{2m}}p_2 &
        \sqrt{\dfrac{\kappa}{2}}u_2
    \end{pmatrix}^T.
\end{equation}
Then we define the inner product as
\begin{multline}
    \bra {\psi_2} \ket{\psi_1} = \dfrac{1}{2}\int_{-\infty}^{\infty} \left[\varepsilon_0 E_2^{*}(x) E_1(x) + \mu_0 H_2^{*}(x)H_1(x) \right] \mathrm{d} x \\+ \dfrac{p_2^{*} p_1}{2m} + \dfrac{\kappa u_2^{*} u_1}{2}.
\end{multline}

At this stage, it becomes clear that defining the system states via the normalized variables $E, H, p$, and $u$ (see Eqs.~\eqref{state1} and~\eqref{state2}) serves a specific physical purpose. With this choice, the inner product of a state with itself directly corresponds to the total energy of the system.

From Eqs.~\eqref{mechanical eqs} and \eqref{maxwell eqs}, one can obtain the Schr\"odinger equation describing the time evolution of the system's state
\begin{equation}
    i \dfrac{\partial}{\partial t} \ket{\psi} = \hat{H} \ket{\psi},
\end{equation}
where the Hamiltonian $\hat{H}$ is given by
\begin{equation}
    \hat{H} = \begin{pmatrix}
        0 & -ic \dfrac{\partial}{\partial x} & - \dfrac{i \sigma}{\sqrt{\varepsilon_0 m}} \delta(x) & 0 \\
        -ic \dfrac{\partial}{\partial x} & 0 & 0 & 0 \\
        \dfrac{i \sigma}{\sqrt{\varepsilon_0 m}} \mathcal{P} & 0 & 0 & -i \omega_0 \\
        0 & 0 & i \omega_0 & 0
    \end{pmatrix}.
\end{equation}
Here, the operator $\mathcal{P}$ is defined as a mapping $L^2(\mathbb{R}) \to \mathbb{C}$ acting as $\mathcal{P} f = \int_{-\infty}^{\infty} f(x) \delta(x) \mathrm{d} x = f(0)$, and $\omega_0 = \sqrt{\kappa/m}$ denotes the natural oscillation frequency.

We should note that, due to the delta function appearing in the Hamiltonian, one must specify the domain on which the Hamiltonian is defined and on which the physical states reside. We define this domain as
\begin{equation}
\begin{aligned}
    D(\hat H)=
    \Bigg\{
    &\ket\Psi=
    \begin{pmatrix}
    e\\ h\\ q\\ r
    \end{pmatrix}
    \in
    L^2(\mathbb{R}) \oplus L^2(\mathbb{R}) \oplus\mathbb{C}\oplus \mathbb{C},
    \\
    & e(0^+)=e(0^-),
    \\
    & h(0^+)-h(0^-)
    =-\sigma\sqrt{\frac{\mu_0}{m}}\,q
    \Bigg\},
\end{aligned}
\end{equation}
where $e = \sqrt{\varepsilon_0/2} E$, $h = \sqrt{\mu_0/2} H$, $q = p/\sqrt{2m}$, and $r = u \sqrt{\kappa/2}$.

In Appendix~\ref{proofofhermiciity}, we explicitly show that the Hamiltonian of the full system $\hat{H}$ is Hermitian. Consequently, its eigenfrequencies are purely real.

\section{Mechanical subsystem}

Let us consider a mechanical subsystem whose states are given by
\begin{equation}
    \ket{\psi_{mech}} = \begin{pmatrix}
        \dfrac{1}{\sqrt{2m}}p &
        \sqrt{\dfrac{\kappa}{2}}u    \end{pmatrix}^T,
\end{equation}
which satisfies the equation
\begin{equation}
\label{mech shordinger}
    i\dfrac{\mathrm{d}}{\mathrm{d} t} \ket{\psi_{mech}} = \hat{H}^0_{mech} \ket{\psi_{mech}} + \begin{pmatrix}
        \dfrac{i\sigma}{\sqrt{2m}}E(0) \\
        0\end{pmatrix},
\end{equation}
where the bare mechanical Hamiltonian $\hat{H}^0_{mech}$ is given by
\begin{equation}
    \hat{H}^0_{mech} = \begin{pmatrix}
        0 & -i\sqrt{\dfrac{\kappa}{m}} \\
         i\sqrt{\dfrac{\kappa}{m}}  & 0
    \end{pmatrix}.
\end{equation}

If we set $\sigma=0$, then the eigenfrequencies of the mechanical subsystem are $\pm \omega_0 = \pm \sqrt{\kappa/m}$. In turn, solving Eq.~\eqref{mech shordinger} in the monochromatic case yields

\begin{equation}
\label{momentum}
    \ket{\psi_{mech}} = \dfrac{\sigma E(0)}{\omega^2 - \omega_0^2}\begin{pmatrix}
        \dfrac{i\omega}{\sqrt{2m}} & \dfrac{-\omega_0}{\sqrt{2m}}
    \end{pmatrix}^T.
\end{equation}
At first glance, Eq.~\eqref{momentum} might be expected to diverge at the resonance frequencies $\pm \omega_0$. However, such a divergence is physically groundless. In the presence of a non-zero surface charge $\sigma \neq 0$, the eigenfrequencies are shifted due to the dipole’s self-action. This self-action arises from the non-zero electric field produced by the dipole at its own position, which can be evaluated as

\begin{equation}
\label{self-field}
    E_{self}(0) = -\dfrac{i}{\varepsilon_0} \int_{-\infty}^{\infty} \hat{G}_{11}(0-x') j(x') \mathrm{d} x',
\end{equation}
where $\hat{G}$ denotes the 2D Green’s function of the electromagnetic subsystem.

Up to this point, the explicit form of the Green’s function for the electromagnetic subsystem has not yet been introduced. In what follows, we derive this Green’s function explicitly and, based on the obtained expression, demonstrate how the eigenfrequencies of the mechanical subsystem are modified.

\section{2D Green's function of electromagnetic subsystem}

In this section, we explicitly derive the 2D Green’s function for the electromagnetic subsystem. The governing equations for this subsystem are expressed as
\begin{equation}
    \begin{pmatrix}
        \omega & \dfrac{i}{\varepsilon_0} \dfrac{\partial}{\partial x} \\
        \dfrac{i}{\mu_0} \dfrac{\partial}{\partial x} & \omega
    \end{pmatrix}
    \begin{pmatrix}
        E(x) \\
        H(x)
    \end{pmatrix} = 
    \begin{pmatrix}
        - \dfrac{i}{\varepsilon_0} j(x) \\
        0
    \end{pmatrix}.
\end{equation}
Therefore, the 2D Green’s function for the electromagnetic subsystem, $\hat{G}(x-x')$, is defined by
\begin{equation}
    \begin{pmatrix}
        \omega & \dfrac{i}{\varepsilon_0} \dfrac{\partial}{\partial x} \\
        \dfrac{i}{\mu_0} \dfrac{\partial}{\partial x} & \omega
    \end{pmatrix}
    \hat{G}(x-x') = 
    \delta(x-x')\mathbbm{1}_{2 \times 2},
\end{equation}
where $\mathbbm{1}_{2 \times 2}$ denotes the $2\times 2$ identity matrix. The Fourier transform of the Green’s function is then given by
\begin{equation}
    \hat{G}(k) = \dfrac{1}{\omega^2 - c^2 k^2} \begin{pmatrix}
        \omega & \dfrac{1}{\varepsilon_0} k \\
        \dfrac{1}{\mu_0} k & \omega
    \end{pmatrix}.
\end{equation}
Transforming back to the coordinate representation yields the Green’s function as
\begin{equation}
\label{Green integral}
    \hat{G}(x-x') = \dfrac{1}{2 \pi} \int_{-\infty}^{\infty} \dfrac{\mathrm{d} k}{\omega^2 - c^2 k^2} \begin{pmatrix}
        \omega & \dfrac{1}{\varepsilon_0} k \\
        \dfrac{1}{\mu_0} k & \omega
    \end{pmatrix} e^{ik(x-x')}.
\end{equation}

The integrand in Eq.~\eqref{Green integral} has two poles at $k_{\pm} = \pm \omega/c$. We introduce the regularization $k_1 = -\omega/c - i\varepsilon$ and $k_2 = \omega/c + i \varepsilon$, where $\varepsilon \to +0$ is an infinitesimal positive parameter. This choice of regularization ensures the radiation condition at infinity.

For $x \neq x'$, closing the integration contour in the complex plane and applying Cauchy’s residue theorem yields
\begin{multline}
\label{Green 1}
    \hat{G}(x-x') = - \dfrac{i}{2c} \begin{pmatrix}
        1 & \dfrac{\mathrm{sign} (x-x')}{\varepsilon_0 c} \\
        \dfrac{\mathrm{sign} (x-x')}{\mu_0 c} & 1
    \end{pmatrix} e^{i \frac{\omega}{c} |x-x'|} \\ \text{ if } x \neq x'.
\end{multline}
When $x=x'$, the Sokhotski–Plemelj theorem allows us to obtain
\begin{multline}
    \hat{G}(0) = \dfrac{1}{2 \pi} \mathrm{P.V.} \int_{-\infty}^{\infty} \dfrac{\mathrm{d} k}{\omega^2 - c^2k^2 } \begin{pmatrix}
        \omega & \dfrac{1}{\varepsilon_0} k \\
        \dfrac{1}{\mu_0} k & \omega
    \end{pmatrix} \\
    -\dfrac{i}{2c} \begin{pmatrix}
        1 & 0 \\
        0 & 1
    \end{pmatrix}.
\end{multline}
It is straightforward to verify that the principal-value integral vanishes. Consequently, the Green’s function at $x=x'$ becomes purely imaginary and reduces to
\begin{equation}
\label{GreenZero}
    \hat{G} (0) = - \dfrac{i}{2 c} \mathbbm{1}_{2 \times 2}.
\end{equation}

\section{Non-Hermitian effective Hamiltonian of mechanical subsystem}

Having obtained the exact expression for the Green’s function, we can now evaluate the dipole’s self-action. According to Eqs.~\eqref{self-field} and~\eqref{GreenZero}, the self-induced component of the electric field at the dipole's position is
\begin{equation}
\label{self field}
    E(0)_{self} = -\dfrac{i \sigma}{\varepsilon_0 m} p \hat{G}_{11}(0) = -\dfrac{\sigma}{2 c \varepsilon_0 m} p.
\end{equation}
Assuming the external electric field consists of two counter-propagating waves, $A_{+} e^{i(\omega/c) x}$ and $A_{-} e^{-i(\omega/c) x}$, its value at the dipole's position is given by
\begin{equation}
\label{ext field}
    E(0)_{ext} = A_{+} + A_{-}.
\end{equation}

Combining Eqs.~\eqref{self field} and~\eqref{ext field} yields the modified equation of time evolution for the mechanical subsystem
\begin{equation}
\label{mech subsystem}
    i\dfrac{\mathrm{d}}{\mathrm{d} t} \ket{\psi_{mech}} = \hat{H}_{mech} \ket{\psi_{mech}} + \hat{D} \ket{s_{+}},
\end{equation}
where
\begin{equation}
    \hat{D} = i \sqrt{2 \gamma c \varepsilon_0} \begin{pmatrix}
        1 & 1 \\
        0 & 0
    \end{pmatrix},
\end{equation}
\begin{equation}
\label{input vector}
    \ket{s_{+}} = \begin{pmatrix}
        A_{+} & A_{-}
    \end{pmatrix}^T,
\end{equation}
and the dressed mechanical Hamiltonian is defined as
\begin{equation}
    \hat{H}_{mech} = \begin{pmatrix}
    -2i \gamma & -i\omega_0 \\
    i\omega_0 & 0
    \end{pmatrix}
\end{equation}
with $\gamma = \sigma^2 |\mathrm{Im}(\hat{G}_{11}(0))|/2\varepsilon_0m$.

Consequently, the eigenfrequencies of the mechanical subsystem become complex due to the self-action and are given by
\begin{equation}
\label{1dipolepoles}
    \omega_{\pm} = -i \gamma \pm \sqrt{\omega_0^2 - \gamma^2}.
\end{equation}
Thus, we have shown that the radiative damping rate $\gamma$ arises from the coupling between the mechanical and electromagnetic subsystems. This coupling is driven by the non-zero charge density $\sigma$, while its dissipative character originates from the purely imaginary value of the 2D electromagnetic Green’s function at the dipole’s position.

\section{Poles of $S$-matrix}

In this section, we demonstrate how the eigenfrequencies of the mechanical subsystem influence the resonances manifesting in the $S$-matrix. Within our model, the electric field can be expressed in terms of the previously derived 2D electromagnetic Green’s function as
\begin{equation}
\label{field through GF}
    E(x) = A_{+}e^{i(\omega/c)x} + A_{-}e^{-i(\omega/c)x} - \dfrac{i}{\varepsilon_0} \int_{-\infty}^{\infty} G_{11}(x-x') j(x') \mathrm{d} x'.
\end{equation}
The amplitudes of the incoming waves can be grouped into the vector $\ket{s_{+}} = \begin{pmatrix}
A_{+} & A_{-}\end{pmatrix}^T$. The $S$-matrix relates these incoming amplitudes $\ket{s_{+}}$ to the outgoing amplitudes $\ket{s_{-}}$ via
\begin{equation}
    \ket{s_{-}} = \hat{S} \ket{s_{+}}.
\end{equation}

To determine the outgoing amplitudes, we evaluate the electric field on both sides of the dipole using Eq.~\eqref{field through GF}. Specifically, for $x<0$, we have
\begin{multline}
    E(x) = A_{+}e^{i(\omega/c)x} + A_{-}e^{-i(\omega/c)x} \\ + \dfrac{i}{c \varepsilon_0} \left[ \hat{D}^T \ket{\psi_\mathrm{mech}} \right]_1 e^{-i(\omega/c)x},
\end{multline}
while for $x>0$, the field reads
\begin{multline}
    E(x) = A_{+}e^{i(\omega/c)x} + A_{-}e^{-i(\omega/c)x} \\ +\dfrac{i}{c \varepsilon_0} \left[ \hat{D}^T \ket{\psi_\mathrm{mech}} \right]_2 e^{i(\omega/c)x}.
\end{multline} 

According to Eq.~\eqref{mech subsystem}, which governs the dynamics of the mechanical subsystem, the monochromatic solution takes the form
\begin{equation}
    \ket{\psi_\mathrm{mech}} = \left(\omega \mathbbm{1}_{2 \times 2} - \hat{H}_\mathrm{mech} \right)^{-1} \hat{D} \ket{s_{+}}.
\end{equation}
Substituting this solution into the expressions for the electric field yields the $S$-matrix of the system:
\begin{equation}
\label{Smatrix}
    \hat{S} = \begin{pmatrix}
        0 & 1 \\
        1 & 0
    \end{pmatrix} \\ - \dfrac{i}{c \varepsilon_0} \left[\hat{D}^T(\hat{H}_\mathrm{mech} - \omega \mathbbm{1}_{2\times2})^{-1} \hat{D} \right].
\end{equation}

Equation~\eqref{Smatrix} demonstrates that the poles of the $S$-matrix are determined by the resolvent of the dressed mechanical Hamiltonian, $(\hat{H}_\mathrm{mech} - \omega \mathbbm{1}_{2\times2})^{-1}$. Hence, the resonances of the $S$-matrix are directly dictated by the eigenfrequencies of the mechanical subsystem. The imaginary part of these poles is intrinsic to the mechanical subsystem, where it is induced by the coupling to the electromagnetic subsystem via the non-zero charge density $\sigma$ and the imaginary part of the Green’s function, $|\mathrm{Im}G_{11}(0)|$.

\begin{figure}
\centering
\includegraphics[width = 1\columnwidth]{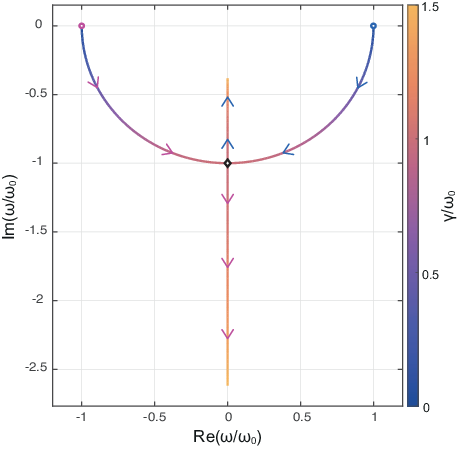}
    \caption{Poles of the $S$-matrix in the complex frequency plane as a function of the normalized damping rate $\gamma/\omega_{0}$. The motion of poles $\omega_{+}$ and $\omega_{-}$ across complex plane is represented by the blue and purple arrows, respectively.}
\label{poles1}
\end{figure}

\begin{figure}
\centering
\includegraphics[width = 1\columnwidth]{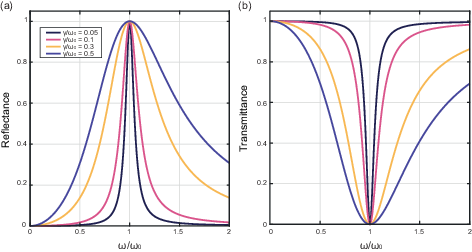}
    \caption{(a) Reflection and (b) transmission of the dipole as functions of the normalized frequency $\omega/\omega_0$. The different curves represent various values of the radiative loss rate $\gamma$ shown in the insets.}
\label{Spectra1}
\end{figure}

The poles of the $S$-matrix (see Eq.~\eqref{1dipolepoles}) as a function of the normalized damping rate $\gamma/\omega_0$ are shown in Fig.~\ref{poles1}. Starting from two isolated points, when reaching $\gamma = \omega_0$, they merge into a single purely imaginary pole and with increasing ratio, $\gamma/\omega_0 \gg 1$, they diverge in different directions on an imaginary axis. The reflection and transmission spectra as functions of the normalized frequency $\omega/\omega_0$ are depicted in Fig.~\ref{Spectra1} for different values of the ratio $\gamma/\omega_0$. In the limit $\gamma \ll \omega_0$, the resonance peaks tend toward a narrow Lorentzian shape, whereas with an increasing damping rate $\gamma$, their profiles become wider and more asymmetric.

\section{Fabry-P\'erot resonances}

Now, using our approach, we analyze a pair of dipoles forming a Fabry-P\'erot resonator. We consider a two-resonance system because a single dipole yields no spectral restructuring when coupled to the continuum. In this configuration, we can formally treat the interaction with the continuum as a perturbation. Counterintuitively, even if this perturbation is arbitrarily weak, it does not merely perturb the two existing modes of the isolated dipoles. Instead, it qualitatively changes the system's spectrum by instantly inducing an additional discrete Fabry-P\'erot series that is fundamentally absent in the single-dipole or uncoupled cases.

Let us consider two dipoles separated by a distance $L$. The mechanical equations describing this system read as follows
\begin{equation}
\label{mech 2 shordinger}
    i\dfrac{\mathrm{d}}{\mathrm{d} t} \ket{\phi_\mathrm{mech}} = \hat{\mathcal{H}}^0_{mech} \ket{\phi_\mathrm{mech}} + \begin{pmatrix}
        \dfrac{i\sigma_1}{\sqrt{2m_1}}E(0) \\
        0 \\
        \dfrac{i\sigma_2}{\sqrt{2m_2}}E(L) \\
        0
        \end{pmatrix},
\end{equation}
where $\hat{\mathcal{H}}^0_\mathrm{mech}$ is the bare mechanical Hamiltonian
\begin{equation}
    \hat{\mathcal{H}}^0_\mathrm{mech} = \begin{pmatrix}
        0 & -i\sqrt{\dfrac{\kappa_1}{m_1}} & 0 & 0\\
        i\sqrt{\dfrac{\kappa_1}{m_1}}  & 0 & 0 & 0 \\
        0 & 0 & 0 & -i\sqrt{\dfrac{\kappa_2}{m_2}}\\
        0 & 0 &  i\sqrt{\dfrac{\kappa_2}{m_2}}  & 0
    \end{pmatrix}
\end{equation}
and the states of the system are defined as
\begin{equation}
    \ket{\phi_\mathrm{mech}} = \begin{pmatrix}
        \dfrac{1}{\sqrt{2m_1}} p_1 & \sqrt{\dfrac{\kappa_1}{2}} u_1 & \dfrac{1}{\sqrt{2m_2}} p_2 & \sqrt{\dfrac{\kappa_2}{2}} u_2
    \end{pmatrix}^T.
\end{equation}

To proceed, the fields $E(0)$ and $E(L)$ must incorporate both the self-action of the dipoles and the cross-action between them. Therefore, we write
\begin{align}
\label{1}
    & E(0) = E_{ext}(0) + E_{self} (0) + E_{adj} (0), \\
    & \notag E(L) = E_{ext}(L) + E_{self} (L) + E_{adj} (L),
\end{align}
where the contribution of the external field is
\begin{align}
\label{2}
    & E(0) = A_{+} + A_{-}, \\
    & \notag E(L) = A_{+}e^{iL\omega/c} + A_{-}e^{-iL\omega/c}, 
\end{align}
the term describing the action from the adjacent dipole is
\begin{align}
\label{3}
    & E_{adj}(0) = - \dfrac{i \sigma_2}{\varepsilon_0 m_2} p_2 G_{11}(L) = - \dfrac{\sigma_2}{2 c \varepsilon_0 m_2} p_2 e^{i L \omega/c}, \\
    & \notag E_{adj}(L) = - \dfrac{i \sigma_1}{\varepsilon_0 m_1} p_1 G_{11}(L) = - \dfrac{\sigma_1}{2 c \varepsilon_0 m_1} p_1 e^{i L \omega/c},
\end{align}
and the self-action term is
\begin{align}
\label{4}
    & E_{self}(0) = -\dfrac{i \sigma_1}{\varepsilon_0 m_1} p_1 G_{11}(0) = -\dfrac{\sigma_1}{2 c \varepsilon_0 m_1} p_1, \\
    & \notag E_{self}(L) = -\dfrac{i \sigma_2}{\varepsilon_0 m_2} p_2 G_{11}(0) = -\dfrac{\sigma_2}{2 c \varepsilon_0 m_2} p_2. 
\end{align}

Combining Eqs.~\eqref{mech 2 shordinger},~\eqref{1},~\eqref{2},~\eqref{3}, and~\eqref{4}, we obtain the effective equation for the mechanical subsystem
\begin{equation}
\label{mech 2 subsystem}
    i\dfrac{\mathrm{d}}{\mathrm{d} t} \ket{\phi_{mech}} = \hat{\mathcal{H}}_{mech} \ket{\phi_{mech}} + \hat{\mathcal{D}} \ket{s_{+}},
\end{equation}
where the input vector $\ket{s_{+}}$ is given in Eq.~\eqref{input vector}, the coupling matrix $\hat{\mathcal{D}}$ is given by
\begin{equation}
    \hat{\mathcal{D}} = i \sqrt{2 c \varepsilon_0} \begin{pmatrix}
        \sqrt{\gamma_1} & \sqrt{\gamma_1} \\
        0 & 0 \\
        \sqrt{\gamma_2}e^{iL\omega/c} & \sqrt{\gamma_2}e^{-iL\omega/c} \\
        0 & 0 
    \end{pmatrix},
\end{equation}
and the dressed mechanical Hamiltonian $\hat{\mathcal{H}}_{mech}$ reads
\begin{equation}
\label{dressed mechanical hamiltonian}
    \hat{\mathcal{H}}_{mech} = \begin{pmatrix}
        -2i \gamma_1 & -i\omega_1 & -2i \sqrt{\gamma_1 \gamma_2} e^{iL\omega/c} & 0\\
        i\omega_1  & 0 & 0 & 0 \\
        -2i \sqrt{\gamma_1 \gamma_2} e^{iL\omega/c} & 0 & -2i \gamma_2 & -i\omega_2\\
        0 & 0 &  i\omega_2  & 0
    \end{pmatrix}.
\end{equation}

Consequently, the eigenfrequencies of the mechanical Fabry-P\'erot subsystem described by Hamiltonian~\eqref{dressed mechanical hamiltonian} can be determined from the following equation
\begin{equation}
    e^{-2iL\omega/c} = r_1(\omega) r_2(\omega),
\end{equation}
where the reflection coefficients of the two isolated dipoles are given by
\begin{equation}
    r_{1,2}(\omega) = \dfrac{-2i \gamma_{1,2} \omega}{\omega^2 + 2i \gamma_{1,2} \omega - \omega_{1,2}^2}.
\end{equation}

Next, we address the scattering problem for the system. In the Fabry-P\'erot configuration, the current density takes the form
\begin{equation}
    j(x) = \dfrac{p_1 \sigma_1}{m_1} \delta(x) + \dfrac{p_2 \sigma_2}{m_2} \delta(x-L),
\end{equation}
which allows the electric field to be expressed as in Eq.~\eqref{field through GF}.

To derive the $S$-matrix, we determine the outgoing amplitudes by evaluating the electric field on both sides of the Fabry-P\'erot system. Specifically, in the region $x < 0$, we find
\begin{multline}
    E(x) = A_{+}e^{i(\omega/c)x} + A_{-}e^{-i(\omega/c)x} \\ + \dfrac{i}{c \varepsilon_0} \left[ \hat{\mathcal{D}}^T \ket{\phi_\mathrm{mech}} \right]_1 e^{-i(\omega/c)x},
\end{multline}
while for $x > L$, the field reads
\begin{multline}
    E(x) = A_{+}e^{i(\omega/c)x} + A_{-}e^{-i(\omega/c)x} \\ +\dfrac{i}{c \varepsilon_0} \left[ \hat{\mathcal{D}}^T \ket{\phi_\mathrm{mech}} \right]_2 e^{i(\omega/c)x}.
\end{multline} 

According to Eq.~\eqref{mech 2 subsystem}, which governs the dynamics of the mechanical subsystem, the monochromatic solution takes the form
\begin{equation}
    \ket{\phi_\mathrm{mech}} = \left(\omega \mathbbm{1}_{4 \times 4} -\hat{\mathcal{H}}_\mathrm{mech} \right)^{-1} \hat{\mathcal{D}} \ket{s_{+}}.
\end{equation}
Substituting this solution into the expressions for the electric field yields the $S$-matrix of the system
\begin{equation}
\label{Smatrix FP}
    \hat{S} = \begin{pmatrix}
        0 & 1 \\
        1 & 0
    \end{pmatrix} \\ - \dfrac{i}{c \varepsilon_0} \left[\hat{\mathcal{D}}^T(\hat{\mathcal{H}}_\mathrm{mech} - \omega \mathbbm{1}_{4\times4})^{-1} \hat{\mathcal{D}} \right].
\end{equation}
Consequently, the resonant poles of the scattering matrix are fully determined by the Fabry-P\'erot spectrum of the mechanical subsystem.

\begin{figure}
\centering
\includegraphics[width = 1\columnwidth]{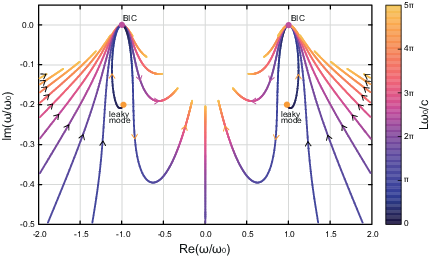}
\caption
{Poles of the $S$-matrix for the system of two identical dipoles having eigenfrequencies $\omega_0$ and damping rates $\gamma_0$, shown as a function of the normalized inter-dipole distance $\omega_0 L/c$. The orange circle marks the initial leaky mode at $L=0$, while the orange arrows indicate its evolution as $L$ increases. Similarly, the purple circle marks the initial BIC at $L=0$, and the purple arrows trace its evolution with increasing $L$. The black arrows indicate the migration of the Fabry-P\'erot resonances, which emerge only for $L\neq 0$.} 
\label{PolesFP}
\end{figure}

Figure~\ref{PolesFP} illustrates the dependence of the $S$-matrix poles for the system of two identical dipoles (with eigenfrequencies $\omega_0$ and damping rates $\gamma_0$) on the normalized inter-dipole distance $\omega_0 L/c$. At $L=0$, the system exhibits two resonances, $\omega_{\pm} = -2i\gamma_0 \pm \sqrt{\omega_0^2 - 4\gamma_0^2}$, with non zero damping rate and two bound states in the continuum (BICs), $\pm \omega_0$. As $L$ increases, these four poles begin to move across the complex plane, while additional Fabry--P\'erot poles emerge from infinity in the lower half-plane. The similar effect of moving of eigenfrequencies across complex plane when changing relative distance in the Fabry--P\'erot system was observed in Ref.~\cite{Dmitriev2019}. Moreover, at specific values of $L$ some modes of Fabry-P\'erot system become BICs due to vanishing imaginary part of their eigenfrequencies.

\begin{figure}
\centering
\includegraphics[width = \columnwidth]{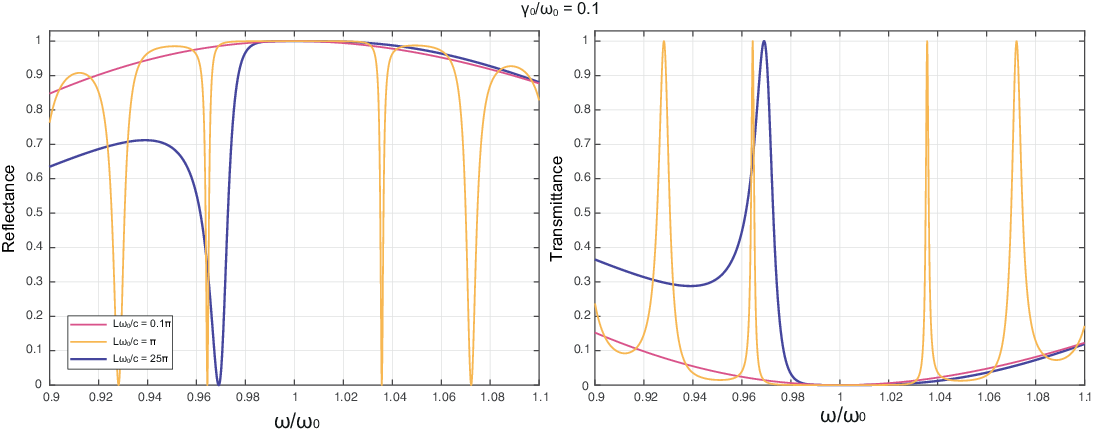}
\caption{(a) Reflection and (b) transmission spectra for the system of two identical dipoles (with eigenfrequencies $\omega_0$ and damping rates $\gamma_0$) as functions of the normalized frequency $\omega/\omega_0$. The ratio $\gamma_0/\omega_0$ is fixed at $0.1$. The different curves correspond to various values of the normalized inter-dipole distance $\omega_0 L/c$.}
\label{SpectraFP}
\end{figure}

The reflection and transmission spectra of the system for a fixed ratio $\gamma_0/\omega_0 = 0.1$ are plotted in Fig.~\ref{SpectraFP} as functions of the normalized frequency $\omega/\omega_0$ for various values of $\omega_0 L/c$. In the limit of small separation (\(\omega_0 L/c = 0.1 \pi\)), we observe a Fano-like resonance profile arising from the interference between a narrow BIC-like mode and a broad resonance associated with the dipoles themselves. At \(\omega_0 L/c = \pi\), the BIC condition is satisfied. However, since this state is a true BIC and therefore decoupled from the radiation continuum, it does not manifest itself in the scattering spectra. Instead, the scattering response at this frequency is governed by the perfect reflection of the individual resonant metasurfaces. An increase in the normalized distance ($\omega_0 L/c = 25 \pi$) clearly demonstrates the emergence of Fabry--P\'erot resonances.

\section{Discussion}

The model developed here provides a clear connection between conservative light–matter dynamics and an effective non-Hermitian description of photonic resonances. In conventional electrodynamics, material degrees of freedom are commonly incorporated into constitutive relations rather than retained as independent dynamical variables. Although this procedure does not necessarily make the electromagnetic problem non-Hermitian, it obscures the partition between material oscillations and the radiation continuum. By retaining both subsystems, our formulation makes this partition explicit and establishes a robust framework in which a Feshbach-type reduction can be performed directly.

A key element of our method is casting the coupled Maxwell and mechanical equations as a first-order evolution equation. In contrast to the conventional second-order wave equation, this approach matches the operator structure of the time-dependent Schr\"odinger equation, cast in a common state space for both field and oscillator variables. To eliminate the electromagnetic subsystem, one utilizes the Green's function of the electromagnetic operator. Since our formulation maintains Maxwell's equations in the first-order form, we first evaluate the Green's function for this operator. The resulting expressions, Eqs.~\eqref{Green 1} and~\eqref{GreenZero}, allow us to express the electromagnetic field via the mechanical variables and substitute it into the equations of motion. This procedure yields the effective mechanical Hamiltonian, which explicitly incorporates both radiative self-action and mutual dipole interactions. While the present model considers one-dimensional propagation, extending the construction of the corresponding Green's function to three spatial dimensions follows an analogous procedure.

The Hermiticity of the full Hamiltonian establishes the conservative character of the interaction: energy transferred from the mechanical oscillator to the electromagnetic field is strictly conserved within the complete system. Radiative decay emerges only in the reduced description, which is obtained by eliminating the electromagnetic degrees of freedom via the outgoing Green's function. In particular, the damping parameter $\gamma$ is determined by the charge-mediated coupling and the local electromagnetic response. Thus, the decay term is rigorously derived from the underlying equations rather than introduced phenomenologically. 
The imaginary part of the reduced resonance frequency describes the energy carried away by radiation, ensuring full consistency with the conservation of total light--matter energy.

The two-dipole system exhibits a remarkable spectral transformation as the separation $L$ deviates from zero. 
At $L=0$, the propagation factor $\exp(i\omega L/c)$ equals unity, and the effective mechanical Hamiltonian yields only a finite set of discrete modes. However, for any nonzero separation (no matter how small) the factor $\exp(i\omega L/c)$ renders the characteristic equation transcendental, instantly giving rise to a countably infinite family of Fabry–P\'erot-like resonances. Thus, a continuous variation of a geometric parameter triggers a singular transition, altering the resonance spectrum from a finite set to a countably infinite one.

This spectral transformation highlights the crucial role of the electromagnetic field in mediating the interaction between the dipoles. Once the dipoles are spatially separated, their coupling incorporates a finite propagation time, the spectral consequences of which remain encoded within the effective mechanical operator even after the field degrees of freedom have been eliminated. How these newly emerged resonance branches formally relate to the degrees of freedom of the complete electromagnetic continuum remains an open question that requires further investigation. From the standpoint of applications, the observed evolution of these emerging branches may provide a promising route toward the realization of higher-order exceptional points.

Finally, the present dipole model should be viewed as a starting point for a broader description of resonant electrodynamics. More complex systems could be represented by a set of material degrees of freedom associated with electric and magnetic multipole moments, coupled through the electromagnetic Green's operator. Preserving the conservative structure of such an enlarged system would then strictly constrain the reduced interactions and radiative decay terms. Establishing the completeness and practical accuracy of these advanced representations remains a promising avenue for future work.

\section{Conclusion}

In conclusion, we have developed a conservative light–matter framework that establishes a rigorous Feshbach projection description for resonant electrodynamics. By explicitly retaining both mechanical and electromagnetic degrees of freedom, we have formulated the coupled dynamics through a strictly Hermitian first-order Hamiltonian, bridging the gap between exact Maxwell's equations and effective mode theories. We have demonstrated that eliminating the electromagnetic subsystem via its outgoing Green’s function yields an effective non-Hermitian mechanical operator, where radiative damping is directly determined by the microscopic coupling and the local field response, rather than introduced phenomenologically. Furthermore, the resulting operator successfully maps the reduced subsystem dynamics onto the resonance poles of the scattering matrix.

We have extended this approach to a system of two spatially separated dipoles, where the formalism inherently incorporates retarded interactions and naturally gives rise to a Fabry–P\'erot-like resonance condition. This example illustrates a profound physical phenomenon: how a finite set of material degrees of freedom acquires a countably infinite resonance structure through its coupling to a propagating continuum.

Ultimately, the framework established in this work opens up a systematic route for applying advanced projection methods and spectral tools from quantum mechanics to complex, resonant photonic systems involving multiple oscillators and higher multipoles in higher spatial dimensions.

\section*{Acknowledgment}
We acknowledge support from the Russian Science Foundation (Grant No. 20-79-10316) and the Belarusian Republican Foundation for Fundamental Research (Project No. F26RNF-106).

\appendix
\section{Proof of Hermicity of full Hamiltonian $\hat{H}$}
\label{proofofhermiciity}

Hermicity of full Hamiltonian $\hat{H}$ would means
\begin{equation}
    \bra{\psi_2}\ket{\hat{H} \psi_1}_{L_2 \oplus L_2 \oplus\mathbb{C} \oplus \mathbb{C}} = \bra{\hat{H}\psi_2} \ket{\psi_1}_{L_2 \oplus L_2 \oplus\mathbb{C} \oplus \mathbb{C}}.
\end{equation}
Meanwhile,
\begin{equation}
    \hat{H} \ket{\psi_1} = \begin{pmatrix}
        -ic \dfrac{\partial h_1}{\partial x} - \dfrac{i \sigma}{\sqrt{\varepsilon_0 m}} \delta(x) q_1 \\
        -ic \dfrac{\partial e_1}{\partial x} \\
        \dfrac{i \sigma}{\sqrt{\varepsilon_0 m}} e_1 (0) - i \omega_0 r_1 \\
        i \omega_0 q_1
    \end{pmatrix}.
\end{equation}
Then we evalute scalar product
\begin{multline}
    \bra{\psi_2} \ket{\hat{H} \psi_1}_{L_2 \oplus L_2 \oplus\mathbb{C} \oplus \mathbb{C}} = \\\int_{-\infty}^{\infty} \left[-ic e^{*}_2 \dfrac{\partial h_1}{\partial x} - \dfrac{i \sigma}{\sqrt{\varepsilon_0 m}} \delta(x) e^{*}_2q_1 - ic h^{*}_2 \dfrac{\partial e_1}{\partial x}\right] \mathrm{d} x\\
    + \dfrac{i \sigma}{\sqrt{\varepsilon_0 m}} q^{*}_2 e_1(0) - i \omega_0 q^{*}_2 r_1 + i\omega_0r^{*}_2q_1 \\= \int_{-\infty}^{\infty} \left[-ic e^{*}_2 \dfrac{\partial h_1}{\partial x} - ic h^{*}_2 \dfrac{\partial e_1}{\partial x}\right] \mathrm{d} x \\- \dfrac{i \sigma}{\sqrt{\varepsilon_0 m}} e^{*}_2(0) q_1  + \dfrac{i \sigma}{\sqrt{\varepsilon_0 m}} q^{*}_2 e_1(0) - i \omega_0 q^{*}_2 r_1 + i\omega_0 r^{*}_2q_1.
\end{multline}

Doing integration by parts and meaning that fields $e_1$, $e_2$, $h_1$, $h_2$ decay at infinity since they belong to $L_2$, we obtain
\begin{multline}
     \bra{\psi_2} \ket{\hat{H} \psi_1}_{L_2 \oplus L_2 \oplus\mathbb{C} \oplus \mathbb{C}} = \\
     \int_{-\infty}^{\infty} \left[ic e_1 \dfrac{\partial h^{*}_2}{\partial x} + ic h_1 \dfrac{\partial e^{*}_2}{\partial x}\right] \mathrm{d} x \\- \dfrac{i \sigma}{\sqrt{\varepsilon_0 m}} e^{*}_2(0) q_1  + \dfrac{i \sigma}{\sqrt{\varepsilon_0 m}} q^{*}_2 e_1(0) - i \omega_0 q^{*}_2 r_1 + i\omega_0 r^{*}_2q_1 \\
     = \bra{\hat{H}\psi_2} \ket{\psi_1}_{L_2 \oplus L_2 \oplus\mathbb{C} \oplus \mathbb{C}}.
\end{multline}
Therefore, $\hat{H}$ is an hermitian operator on $L_2 \oplus L_2 \oplus\mathbb{C} \oplus \mathbb{C}$ space.

\bibliographystyle{apsrev4-2}
\bibliography{cas-refs}

\end{document}